\documentclass[aps,pra,amssymb,superscriptaddress,10pt,tightenlines,twocolumn]{revtex4-1}

\usepackage{graphicx}
\usepackage{dcolumn}
\usepackage{bm}
\usepackage{braket}
\usepackage{amsmath}
\usepackage{footnote}
\usepackage[utf8]{inputenc}
\usepackage[T1]{fontenc}
\usepackage{ragged2e}
\usepackage{subfigure}
\usepackage[dvipsnames]{xcolor}

\newcommand{\ii}{\mathrm{i}}
\newcommand{\e}{\mathrm{e}}

\newcommand{\Res}{\operatorname{Res}}

\begin{document}


\title{Photon-emitter dressed states in a closed waveguide}

\author{Davide Lonigro}
\affiliation{Dipartimento di Fisica and MECENAS, Universit\`{a} di Bari, I-70126 Bari, Italy}
\affiliation{INFN, Sezione di Bari, I-70126 Bari, Italy}

\author{Paolo Facchi}
\affiliation{Dipartimento di Fisica and MECENAS, Universit\`{a} di Bari, I-70126 Bari, Italy}
\affiliation{INFN, Sezione di Bari, I-70126 Bari, Italy}

\author{Andrew D. Greentree}
\affiliation{Australian Research Council Centre of Excellence for Nanoscale Biophotonics, School of Science, RMIT University, VIC 3001 Melbourne, Australia}

\author{\\ Saverio Pascazio}
\affiliation{Dipartimento di Fisica and MECENAS, Universit\`{a} di Bari, I-70126 Bari, Italy}
\affiliation{INFN, Sezione di Bari, I-70126 Bari, Italy}

\author{Francesco V. Pepe}
\affiliation{Dipartimento di Fisica and MECENAS, Universit\`{a} di Bari, I-70126 Bari, Italy}
\affiliation{INFN, Sezione di Bari, I-70126 Bari, Italy}

\author{Domenico Pomarico}
\affiliation{Struttura Semplice Dipartimentale di Fisica Sanitaria, I.R.C.C.S. Istituto Tumori ``Giovanni Paolo II'', I-70124 Bari, Italy}

\date{\today}

\begin{abstract}
We study a system made up of one or two two-level quantum emitters, coupled to a single transverse mode of a closed waveguide, in which photon wavenumbers and frequencies are discretized, and characterize the stable states in which one excitation is steadily shared between the field and the emitters. We unearth finite-size effects in the field-emitter interactions and identify a family of dressed states, that represent the forerunners of bound states in the continuum in the limit of an infinite waveguide. We finally consider the potential interest of such states for applications in the field of quantum information.
\end{abstract}

\pacs{Valid PACS appear here}
\maketitle

\textit{Introduction.---}
One-dimensional and quasi-one-dimensional systems are attracting increasing interest~\cite{onedim_review}, due both to the fundamental and technological relevance of their phenomenology and to the wide range of robust and versatile available experimental platforms, in which dimensional reduction can be efficiently implemented. Among such platforms, it is worth mentioning those based on optical fibers~\cite{onedim3,onedim4}, cold atoms~\cite{focused1,focused2,focused3}, circuit QED~\cite{qdot_review,onedim5,onedim6,mirror1,mirror2,atomrefl1,leo5}, photonic crystals~\cite{kimble1,kimble2,onedim1,onedim2,ck}, and quantum dots in photonic nanowires~\cite{semiinfinite1,semiinfinite2}. In these systems, light propagates in a quasi-one-dimensional geometry with different energy dispersion relations and emitter-photon interaction form factors, determining dimension-dependent features that heavily affect dynamics, decay and propagation~\cite{cirac1,cirac2}. 

Recent work has been devoted to understand the physics of light-matter coupling in waveguides, systems made up of either single quantum emitters~\cite{focused1,mirror2,Lombardo,boundstates1,lalumiere,threelevel}, or sets of two~\cite{refereeA1,refereeA2,PRA2016,oscillators,baranger,baranger2013,NJP,yudson2014,laakso,pichler,Fedorov1,Fedorov2,Calajo,PRA2018,Sinha} or even more emitters~\cite{pichler2,bello,bernien,dong,fang,fang14,goban,ck,gu,guimond,lalumiere,lodahl,paulisch,ramos14,ramos,cirac1,boundstates1,tsoi,yudsonPLA,yudson2008,calajo15,Kockum,Dinc,PRA2019}. In the latter cases, dynamics is deeply influenced by photon-mediated quantum correlations between the emitters, leading to collective phenomena, such as the emergence of superradiant and subradiant states and correlated photon emission.
Such effects generalize Dicke-states and related phenomena, occurring when the optical wavelength is much larger than the typical interatomic distances~\cite{Dicke,SRreview,Kaiser1,Kaiser2}, towards wavelengths that are comparable with the interatomic distance, bringing to light a number of interesting quantum resonance effects.    

In this Letter, we characterize the stable states of a system made up of one or two quantum emitters, coupled to a single transverse mode of a closed waveguide. The analysis is performed in the limit in which only the length of the guide is relevant, the other important physical features being independent of its specific geometry. Unlike in the case of unbounded waveguides, the system admits an infinite number of stable states, regardless of the features of the coupling. However, their characterization highlights features that are specific to different classes of bound states, which can be identified in many cases as forerunners of the bound and quasi-bound (unstable) states in an unbounded geometry. The results are interesting both in view of determining finite-size corrections to the effects identified in the physics of indefinite waveguides, and to identify novel phenomenology that can lead to practical applications. We will show how the properties of a specific class of bound states, related to the wavelength of emitted light resonating with the interatomic distances, suggest the implementation of a qubit in this system.

\begin{figure}
    \centering
	\includegraphics[width=0.22\textwidth]{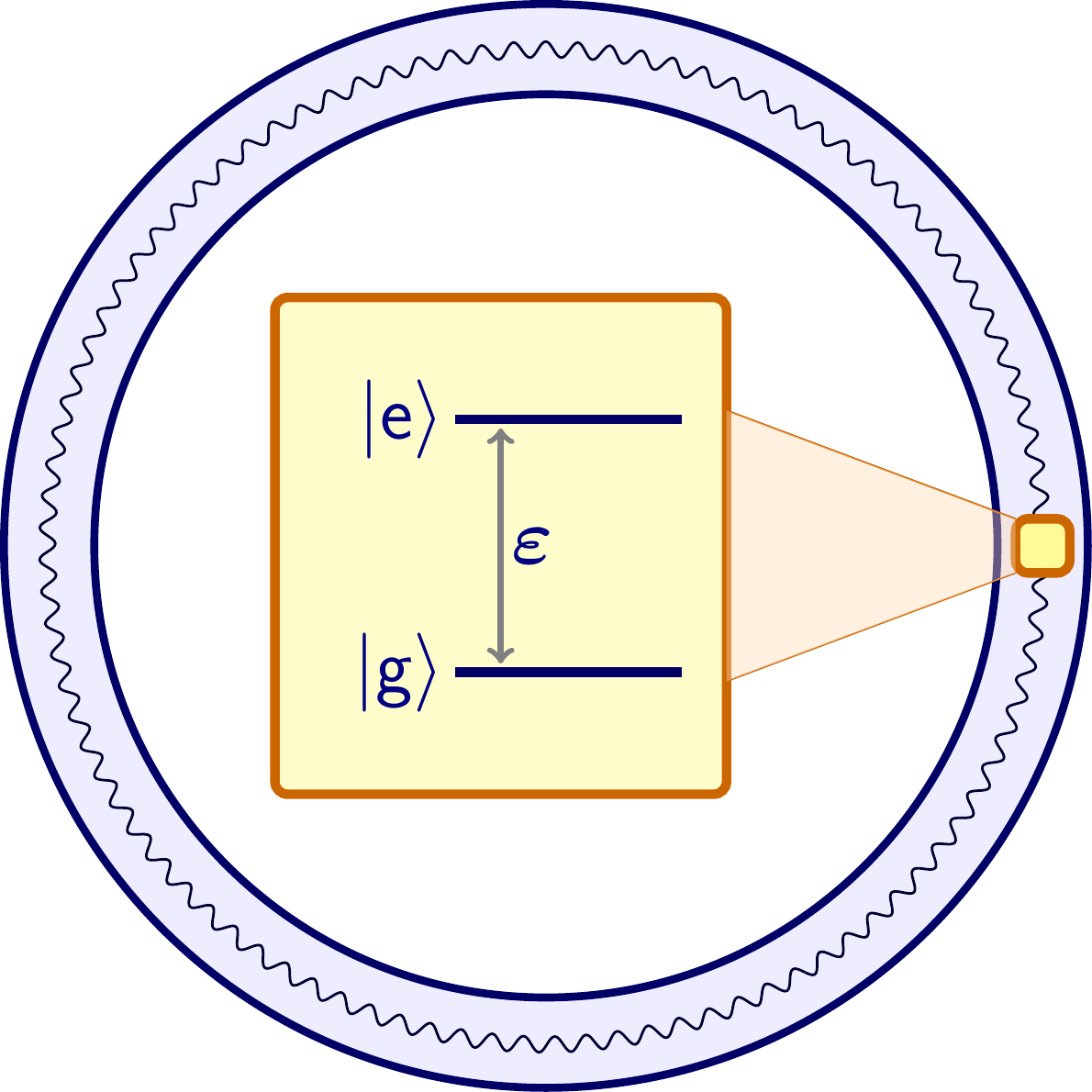} \hspace{0.5cm}
	\includegraphics[width=0.2\textwidth]{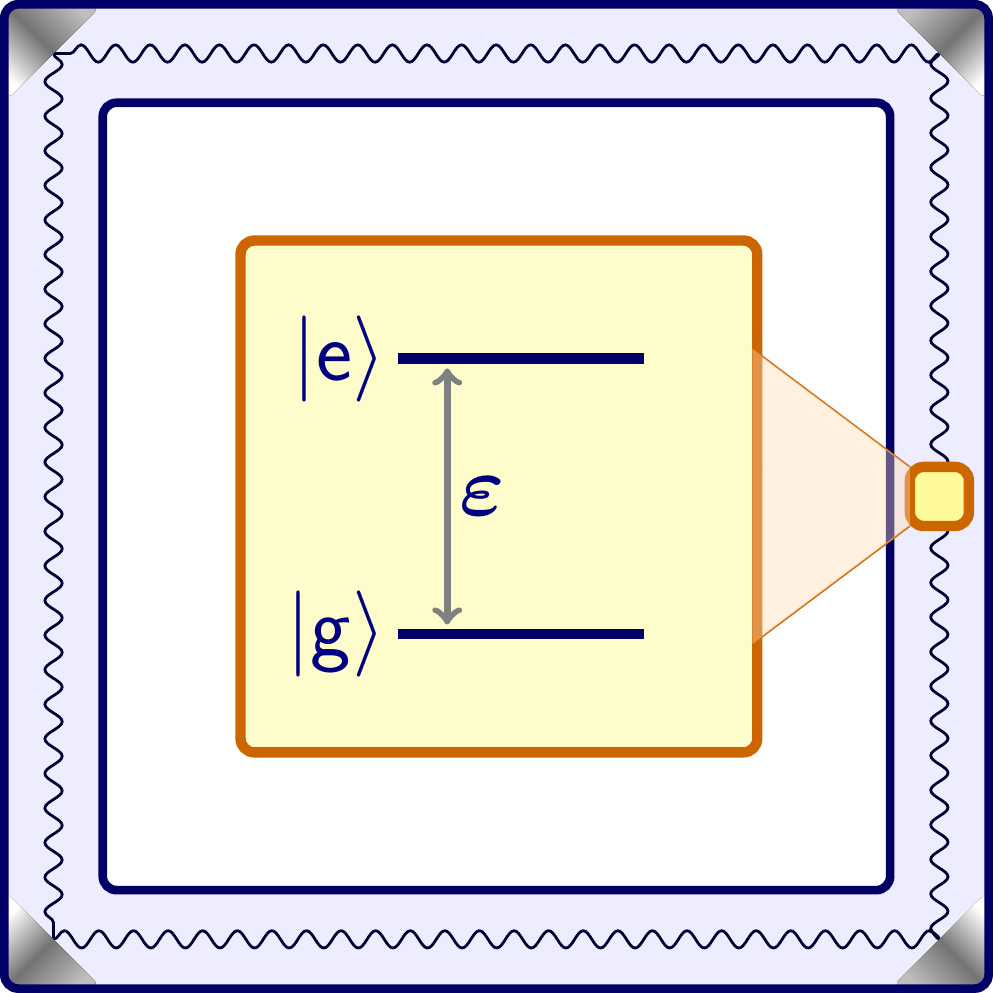}
	\caption{Single two-level emitter coupled to a transverse mode of a closed waveguide, in a circular (left) and square (right) geometry.
	In our analysis the specific geometry of the ring is immaterial,  only its total length $L$ being relevant.}
	\label{fig:wguide}
\end{figure}

\textit{Single emitter.---}
We first consider one quantum emitter in a closed waveguide of length $L$, as represented in Fig.~\ref{fig:wguide}. In the following analysis, we will assume that effects related to the specific geometry of the waveguide are negligible, though they can play a relevant role in the practical realization of the system. The emitter is modeled as a two-level system, with its ground  $\ket{g}$ and excited $\ket{e}$ states separated by the excitation energy $\varepsilon$, and  effectively coupled to a single transverse mode propagating in the waveguide. 
Since the field is also longitudinally confined, the allowed photon wavenumbers can take the discrete values $q_k=2\pi k/L$, with $k\in\mathbb{Z}$, corresponding to the spectrum
\begin{equation}
\label{ke}
	\omega_k = \sqrt{\left(\frac{2\pi k}{L}\right)^2+m^2} \, ,
\end{equation}
where we consider for simplicity a unit system in which the speed of light in the waveguide is $v=1$. In Eq.~\eqref{ke}, the photon effective  mass $m$ is inversely proportional to the transverse size of the guide~\cite{PRA2016}.
The free part of the Hamiltonian thus reads
\begin{equation}
\label{freehamiltonian}
H_0  =   \varepsilon \,\sigma^+ \sigma^-  + \sum_{k=-\infty}^{\infty} \omega_k b^{\dagger}_k b_k ,
\end{equation} 
where $\sigma^+ = (\sigma^-)^\dag = \ket{e}\!\bra{g}$ are the atom ladder operators, and $b_k$ the photon longitudinal mode operators, satisfying the canonical commutation relations $[b_k,b_{k'}]=0$ and $[b_k,b_{k'}^{\dagger}]=\delta_{kk'}$. The interaction Hamiltonian in the electric dipole atom-field coupling and rotating-wave approximation reads
\begin{equation}\label{eq:hint}
H_{\mathrm{int}}=\sum_{k=-\infty}^{\infty} F_k \left(\sigma^+ b_k +  \sigma^- b^\dagger_k\right),
\end{equation}
where the interaction form factor is
\begin{equation}\label{eq:form}
	F_k=\sqrt{\frac{\gamma}{L \omega_k}},
\end{equation}
with $\gamma>0$ a constant with the dimensions of squared energy. 

The rotating-wave form of the interaction~\eqref{eq:hint} allows diagonalization of the Hamiltonian in sectors with fixed number of excitations $\mathcal{N}=\ket{e}\bra{e}+\sum_k b^{\dagger}_k b_k$. In the $\mathcal{N}=1$ sector, the  state of the system reads
\begin{equation}
\label{eq:state}
	\ket{\Psi} = a\ket{e}\otimes\ket{\mathrm{vac}}+\ket{g}\otimes\sum_{k}\xi_k b^\dagger_k \ket{\mathrm{vac}}, 
\end{equation}	
where $\ket{\mathrm{vac}}$ is the field vacuum state and $|a|^2+\sum_{k}|\xi_k|^2=1$. Assume that the atom is placed at $x=0$; 
the photon wavefunction in the position representation $\xi(x)$, with $x\in[-L/2,L/2]$, is given by the Fourier series 
\begin{equation}
	\xi(x)=\frac{2\pi}{L}\sum_k\xi_k \e^{\frac{2\pi \ii kx}{L}}.
\end{equation}
Here, we are interested in the eigenstates of the system, in particular those in which the atomic excitation plays a relevant role. The free Hamiltonian $H_0$ admits two types of eigenstates in the one-excitation sector:  excited atom and no photons $\ket{e}\otimes\ket{\mathrm{vac}}$, with energy $\varepsilon$, or atom in the ground state and a single-photon  $\ket{g}\otimes b_k^{\dagger} \ket{\mathrm{vac}}$, with energy $\omega_k \geq m$.

\begin{figure}[t]
    \centering
	\subfigure[]{\includegraphics[height=0.145\textwidth]{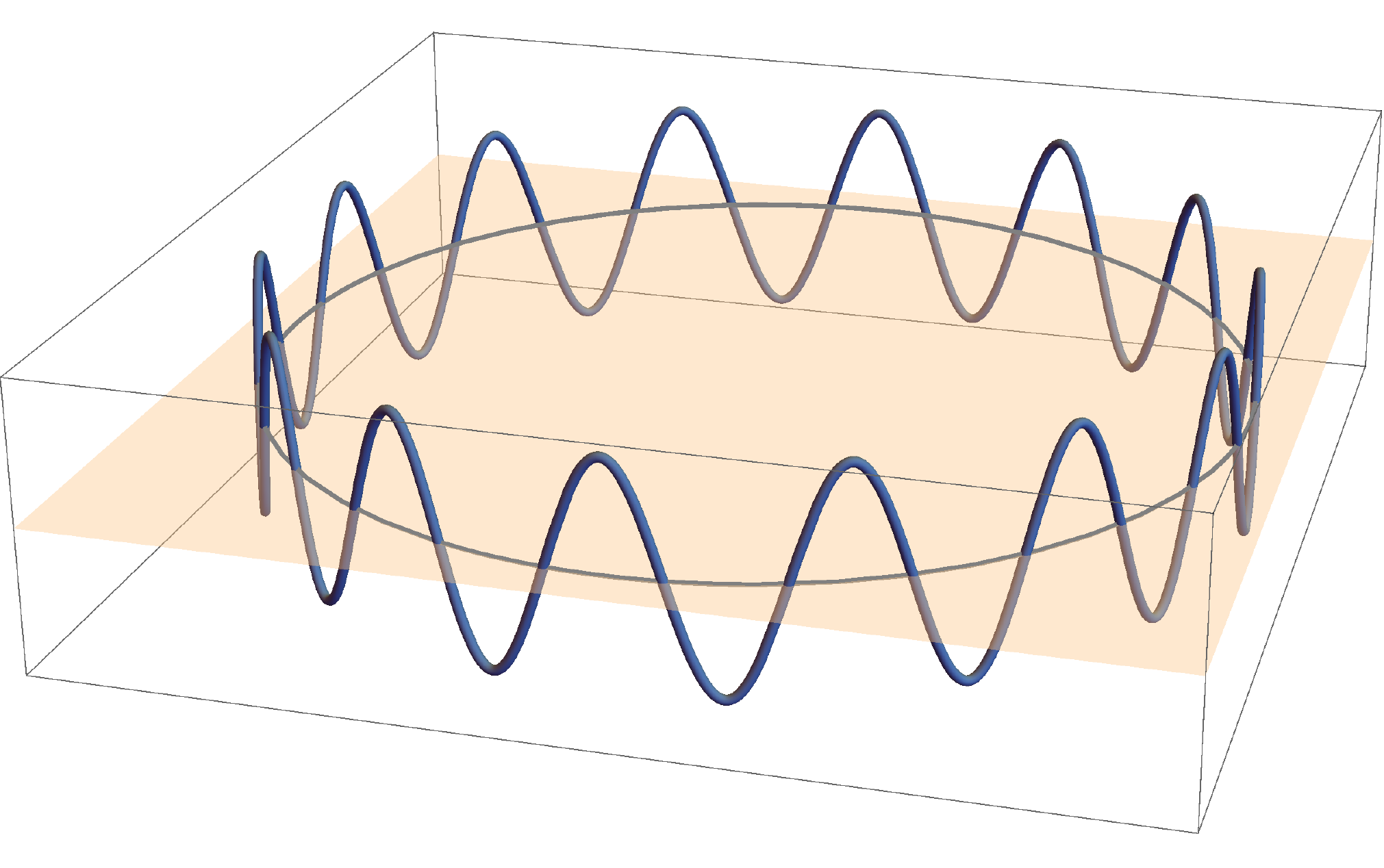}}
    \subfigure[]{\includegraphics[height=0.145\textwidth]{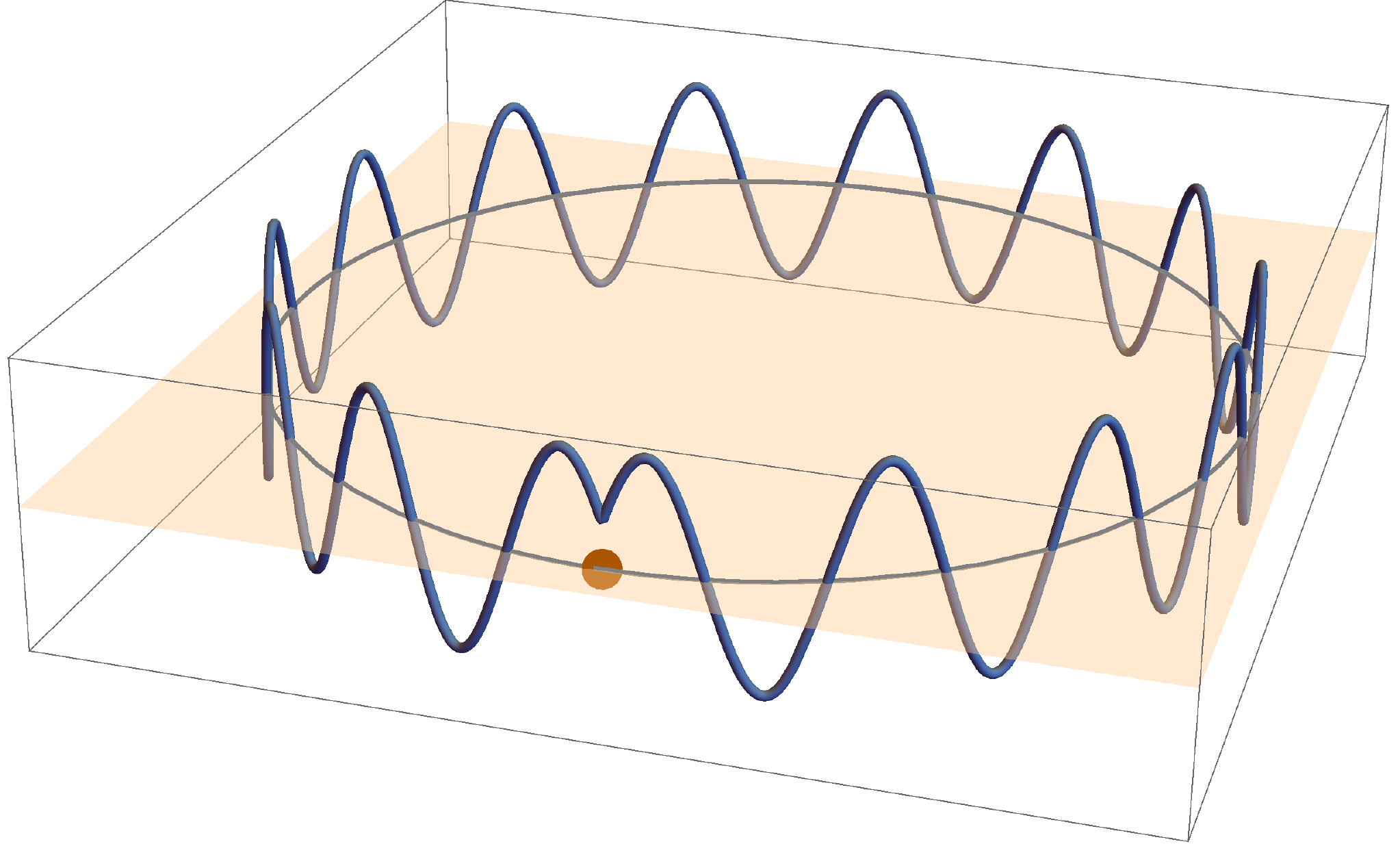}}
\caption{Spatial photon amplitude (blue curve) (a) for an eigenstate of the free Hamiltonian $H_0$ with $\xi_k=\xi_{-k}$, and (b) for an eigenstate of the total Hamiltonian $H=H_0+H_{\mathrm{int}}$. In the latter case, the first derivative of the amplitude has a jump in correspondence of the emitter (red dot): its value increases as the amplitude $a$ of the atomic excitation becomes more relevant.}\label{fig:amplitude}
\end{figure}

When the coupling is switched on, the bound states are modified and the excitation is generally coherently shared between atom and field. For eigenstates of $H=H_0+H_{\mathrm{int}}$ with a nontrivial  atomic component, $a\neq 0$, the eigenenergies $E$ can be determined by solving the equation
\begin{equation}
		E - \varepsilon - \Sigma(E) = 0,\label{eq:eigeq}
\end{equation}
where $\Sigma(E)$ is the self-energy function of the atomic excited state
		$\Sigma(E)= \sum_k F_k^2/(E-\omega_k)$~\cite{cohentannoudji}.
The coefficients~(\ref{eq:state}) are determined by the eigenvalue equation and the normalization condition:
\begin{align}
		a & =\left[ 1 - \Sigma'(E) \right]^{-\frac{1}{2}} , \label{eq:retain} \\
		\xi_k & = \frac{F_k}{E-\omega_k} \left[ 1 - \Sigma'(E) \right]^{-\frac{1}{2}} . \label{eq:boson}
\end{align}
For the form factor~\eqref{eq:form} one gets the analytic expression
\begin{equation}\label{eq:self}
	\Sigma(E)= \gamma\left(\frac{\cot (q(E) L /2)}{q(E)} \theta(E)+ \beta_0(E)\right) ,
\end{equation}
with $q(E)= \sqrt{E^2-m^2}$ and  $\theta$ being the Heaviside step function. 
Here $\beta_0(E)$ is a real-valued function bounded by $|\beta_0(E)|\leq \coth\left(\frac{mL}{2}\right) / (\pi m)$, while the first term in $\Sigma(E)$ diverges at  $E=\omega_k$, for all $k$.  Therefore, the eigenvalue equation~\eqref{eq:eigeq} always admits exactly one solution $E_k$ in each interval $(\omega_k,\omega_{k+1})$, so that the eigenvalues alternate with the photon frequencies: $\omega_k < E_k < \omega_{k+1} < E_{k+1}$ (see Appendix).

\begin{figure}
	\includegraphics[width=0.45\textwidth]{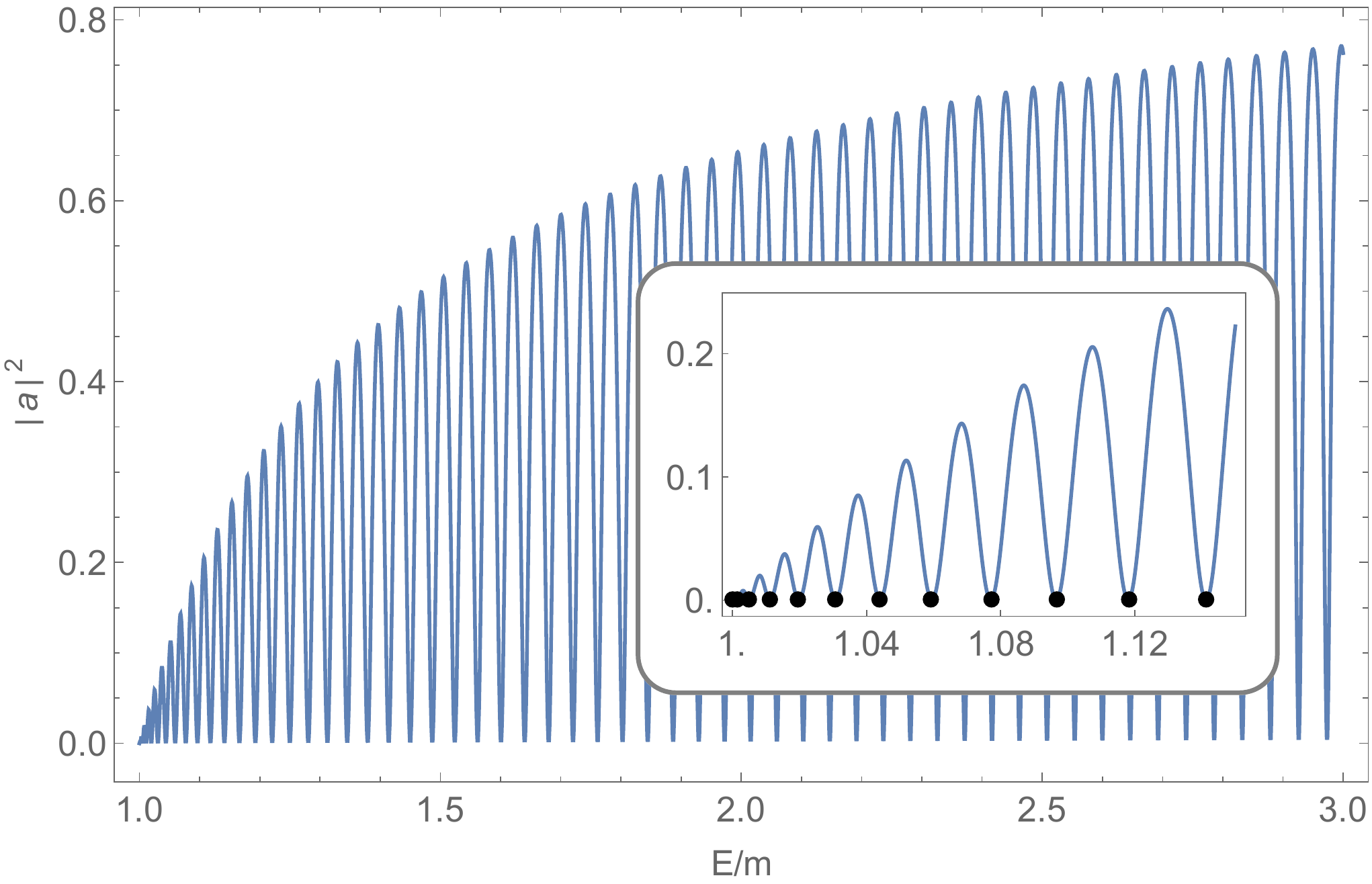}
	\caption{Emitter excitation probability $|a|^2$, as in Eq.~\eqref{eq:retain}, in an eigenstate with energy $E$, for $mL=40\pi$ and $\gamma/m^2=10^{-4}$. The probability that the emitter is in its excited state vanishes at the  photon frequencies $\omega_{k}$, with $k\in \mathbb{Z}$.	
}
	\label{fig:exc}
\end{figure}

 If $\omega_{\ell}<\varepsilon<\omega_{\ell+1}$, we can outline the following typical features of the eigenstates in the perturbative regime: eigenvalues $E_k\neq E_{\ell}$  correspond to states that are generally dominated by a symmetric combination of photon excitations with opposite momenta, whose energy is, respectively, slightly smaller than $\omega_{k+1}$ (for $E_k<E_{\ell}$) or slightly larger than $\omega_k$ (for $E_k>E_{\ell}$); the eigenvalue $E_{\ell}$ is close to $\varepsilon$, up to a correction of $O(\gamma)$, and corresponds to a state with a dominant atomic excitation. A value of $\varepsilon$ very close to a photon frequency $\omega_\ell$ generates two eigenstates, with energies above and below $\omega_\ell$, in which the photon and the atomic excitation are strongly hybridized. By increasing $\gamma$, the eigenvalues $E_k$ migrate towards the center of the intervals $(\omega_k,\omega_{k+1})$, indicating a stronger superposition between atomic and photonic excitations, with a significant involvement of photons with different wavenumbers. Notice that, in any regime, the parameters can be tuned to select a given value of the emitter excitation probability, as summarized in Fig.\ \ref{fig:exc}.

Finally, it is interesting to study what happens in the  limit of a large ring. By increasing the value of $L$, energy eigenvalues  become more and more dense, and the value $|a|^2$ of the atomic excitation probability decreases. In a semiclassical picture, for $L\to\infty$ the photon emitted by the atom will take an infinite time to return to the emitter, which will have released a larger part of its excitation in the meanwhile. None of the eigenstates survives the limit $L\to\infty$, as expected, since the system becomes locally equivalent to an infinite-length linear waveguide, in which no bound state with energy $E>m$ emerges in the single-emitter case, though stable excited states of two or more emitters are possible. 

\textit{Two emitters.---}
We now consider the case in which two identical emitters, with equal excitation energy $\varepsilon$ and positions $x_1$ and $x_2$ at a distance $d$, are coupled to a transverse waveguide mode. Their ground and excited states will be denoted by $\ket{g_\alpha}$ and $\ket{e_\alpha}$, with $\alpha=1,2$, respectively.
The two emitters interact with each other only by photon exchange, and no direct coupling is assumed. The interaction Hamiltonian, which generalizes~\eqref{eq:hint}, reads
\begin{equation}
	H_{\mathrm{int}}=\! \sum_{k=-\infty}^{\infty}  \sum_{\alpha=1,2}F_k \left( \e^{\frac{2\pi \ii k x_\alpha}{L} } \sigma^+_\alpha b_k + \e^{\frac{-2\pi \ii k x_\alpha}{L} } \sigma^-_\alpha b_k^\dag  \right),
\end{equation}
and the system state in the one-excitation sector is
\begin{equation}\label{eq:statetwo}
\ket{\Psi} = \sum_{\alpha=1,2} a_\alpha \sigma^+_\alpha \ket{G} \otimes\ket{\mathrm{vac}} + \ket{G}\otimes \sum_k \xi_k b_k^\dagger\ket{\mathrm{vac}},
\end{equation}
with $\ket{G}=\ket{g_1}\otimes\ket{g_2}$. The bound states of the coupled system represent a generalization of the single-emitter case. The eigenvalue $E$ corresponding to an eigenstate with finite atomic excitation amplitude obeys the equation
\begin{equation}
		\det\left[(E-\varepsilon)\openone-\Sigma(E)\right]=0,
\end{equation}
where the self-energy is now a $2\times 2$ matrix, and $\openone$ is the identity matrix. The atom amplitude vector $\bm{a}=(a_1,a_2)$ characterizing the corresponding eigenstate satisfies
\begin{equation}\label{eq:eigeqvec}
\left[(E - \varepsilon) \openone -\Sigma(E)\right] \bm{a}=0,
\end{equation}
which fixes $a_1$ and $a_2$ up to a global multiplication constant (notice that, generally, $\left\| \bm{a} \right\|< 1$), while the photon amplitudes are determined by a straightforward generalization of~\eqref{eq:boson}. The values of $a_j$ and $\xi_k$ are eventually fixed, up to an overall phase factor, by state normalization. Also in this case, one can derive an analytical form of the self-energy,
\begin{equation}
	\Sigma_{j\ell}(E)= \frac{\gamma}{q(E)} A_{j\ell}\bigl(q(E)\bigr) \theta(E)+\gamma \beta_{j-\ell}(E),
\end{equation} 
where $q(E)= \sqrt{E^2-m^2}$,
\begin{equation}
	A_{j\ell}(q)=\cot\left( \frac{ q L}{2} \right) \,\cos\left((j-\ell)q d\right)+\sin\left(|j-\ell|q d\right),
\end{equation}
$\beta_0$ is the same function that appears in Eq.~\eqref{eq:self}, and $\beta_1=\beta_{-1}$ is a function 
suppressed like
$|\beta_1(E)| \leq [\e^{-m(L-d)}+\e^{-md}]/(\pi m).$
By neglecting the latter contribution for $md, m(L-d)\gg1$, the energies of the bound states are determined by the solutions of the following equations
\begin{equation}\label{eq:epsilontwo}
E \simeq \varepsilon + \frac{\gamma}{q(E)}\chi^{(n)}\bigl(q(E)\bigr) + \gamma \beta_0(E) ,
\end{equation}
where $\chi^{(n)}(q)$, with $n=1,2$, are the eigenvalues, possibly coincident, of the matrix $A(q)$. As in the single-emitter case, the right-hand side diverges at each $\omega_k$, providing a set of energy pairs $\{E^{(1)}_k,E^{(2)}_k\}$ in each interval $(\omega_k,\omega_{k+1})$. The corresponding values the ratio $a_2/a_1$ is determined by the eigenvector of $A$ corresponding to the specific eigenvalue $\chi^{(n)}$. 

\textit{Resonant states.---} The bound state energy can take any value $E>m$, except any of the $\omega_k$'s, depending on $\varepsilon$, $\gamma$ and $m$. A case of particular interest, which is a forerunner of effects observed in an infinite waveguide, is represented by eigenvalues that \textit{resonate} with the distance $d$ between the emitters, namely 
\begin{equation}
	E = E_{\nu} = \sqrt{\left(\frac{\nu\pi}{d}\right)^2 + m^2}, \qquad \nu=1,2,\dots,
\end{equation}
for which an integer number of half-wavelengths $\pi/q(E_\nu)=d/\nu$ (generally not coinciding with any of the half-wavelengths $\pi/q_k=L/(2k)$ of the photon in the ring) separate the two emitters. For this class of stable states, the atomic excitation is determined, up to $O(\e^{-md})$, by the eigenstates of the matrix
\begin{equation}\label{eq:matrix}
	A \left(\frac{\nu\pi}{d}\right)=\cot\left(\frac{\pi\nu L }{2d}\right) \left(\begin{matrix}
	1&(-1)^\nu\\
	(-1)^\nu&1
	\end{matrix}\right) .
\end{equation}
The eigenvalues are $\chi^{(1)}=0$, corresponding to a solution with $a_2/a_1=(-1)^{\nu+1}$, occurring for values of atomic excitation energy satisfying $\varepsilon=E_{\nu} +\beta_0(E_{\nu})$, and $\chi^{(2)}=  2\cot\left(\pi \nu L /2 d\right)$, related to an eigenstates with opposite atomic excitation parity, $a_2/a_1=(-1)^{\nu}$, occurring when $\varepsilon=E_\nu+\beta_0(E_\nu)- 2\gamma d \cot\left(\pi\nu L /2d\right)/ (\nu\pi)$. It is evident that, while the former energy eigenvalue is practically independent of $L$, the latter wildly oscillates as $L$ increases. Indeed, one of the states converges to a true bound state on the infinite waveguide, in which the atomic excitations are combined in such a way to confine the field between the emitter, while the orthogonal combination becomes part of an unstable state, characterized by twice a decay rate of an isolated atom~\cite{PRA2016}. 

\begin{figure}
    \centering
	\subfigure[Atomic excitation amplitudes: $a_1+a_2=0$]{\includegraphics[height=0.135\textwidth]{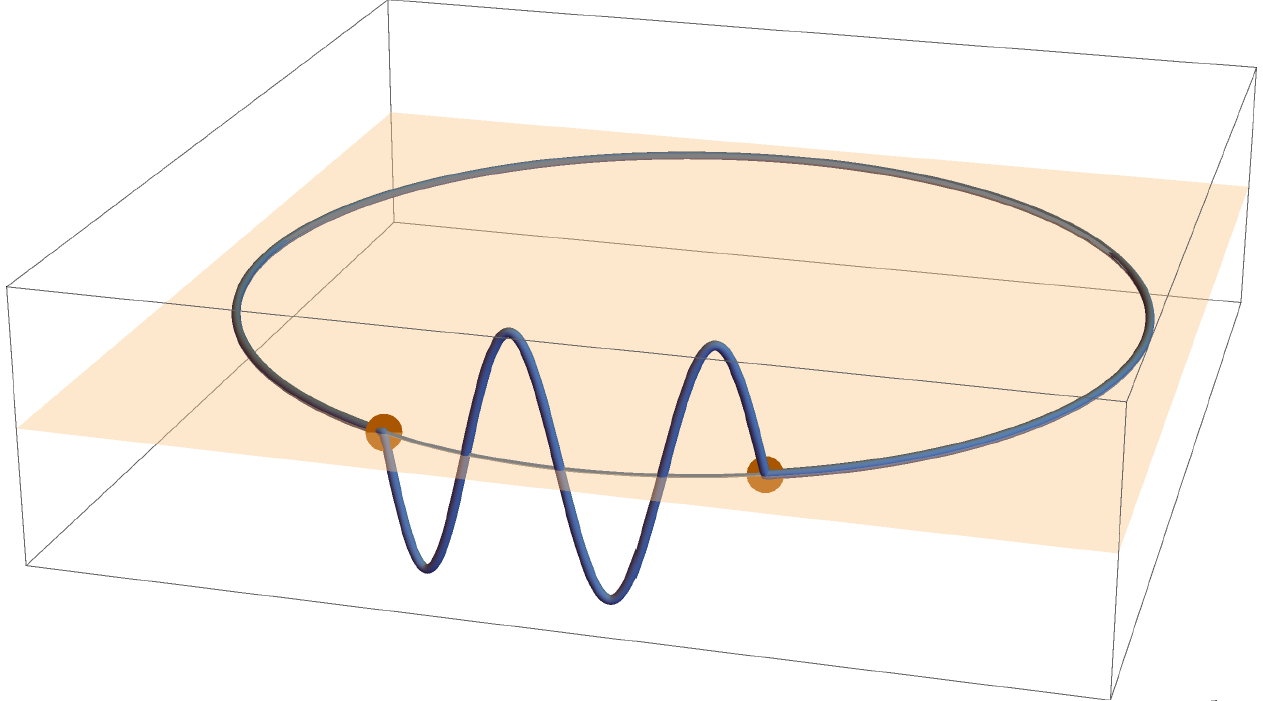}}
	\subfigure[Atomic excitation amplitudes: $a_1-a_2=0$]{\includegraphics[height=0.135\textwidth]{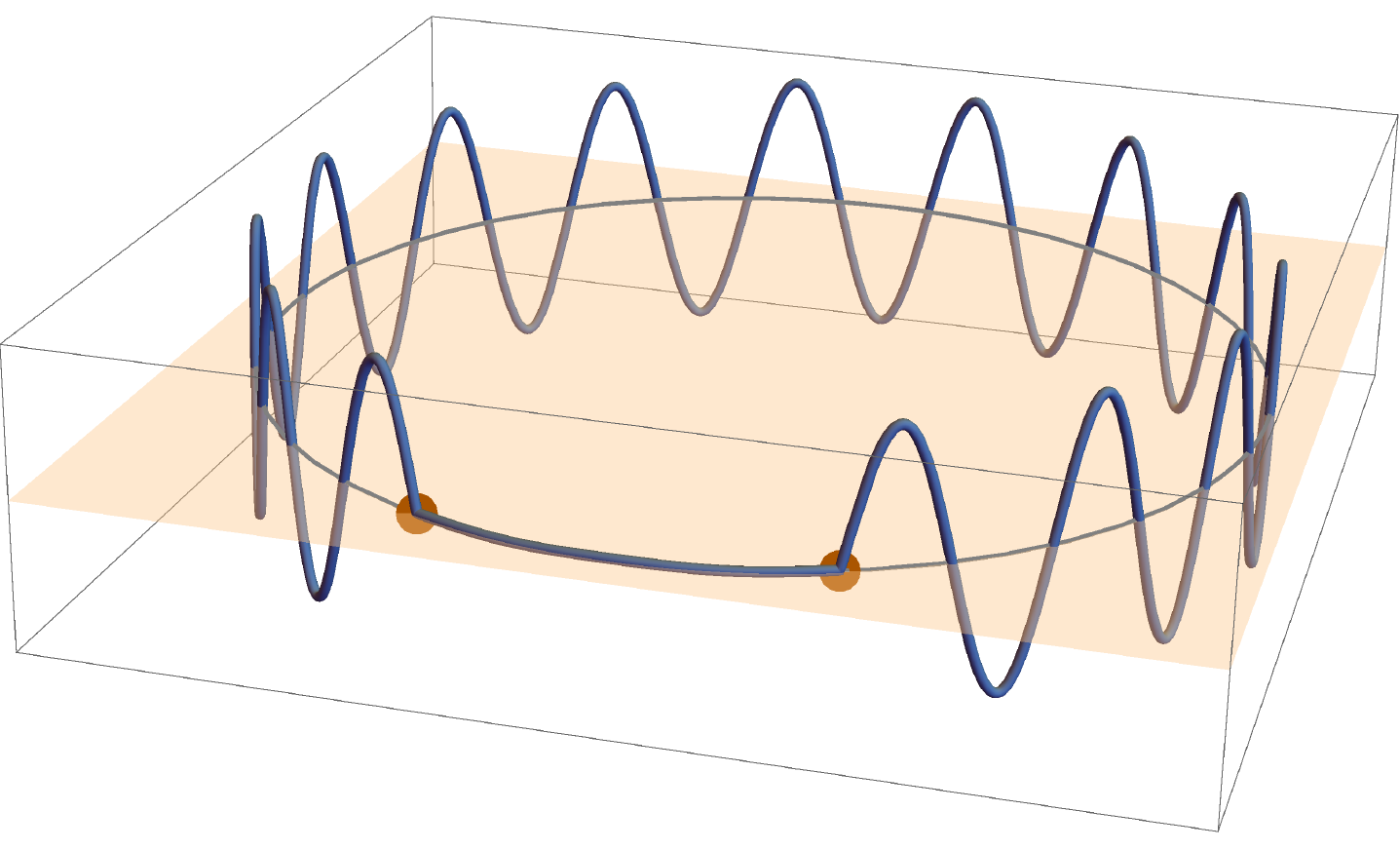}}
	\caption{Photon amplitudes in two resonant bound states occurring in the same system. An integer number of half-wavelength, corresponding to approximately the same energy, can be accommodated in the shortest (of length $d$) and longest (of length $L - d$) path connecting the two emitters.}
	\label{fig:2states}
\end{figure}

Given the value $d$ of the distance, a complementary set of resonant bound states exists, besides the ones discussed above, corresponding to the energies
\begin{equation}
	E = \tilde{E}_{\nu} = \sqrt{\left(\frac{\nu\pi}{L - d}\right)^2 + m^2}, \qquad \nu=1,2,\dots 
\end{equation}
These states are characterized by the fact that the  region between the emitters, of length $L-d$, is (up to $O(\e^{-md})$) a multiple of the half-wavelength associated with the eigenvalue. Their properties are obtained by merely replacing $d$ with $L-d$ in the above equations. However, none of these eigenstates survive as a bound state in the $L\to\infty$ limit.

An interesting situation occurs in the case, depicted in Fig.~\ref{fig:2states}, when the energy $E_{\nu}$ of a resonant state with nonvanishing field in the inner region between the emitters approximately coincides with the energy $\tilde{E}_{{\nu}'}$ of a resonant state with the field confined in the outer region of length $L-d$ and with opposite atomic parity, for some pair $(\nu,\nu')$ of integers. Clearly, this case can occur only when $d$ and $L-d$ are at least approximately commensurable. The relevance of this case lies in the fact that these two states, quasi degenerate and orthogonal to each other, represent the pair $\{E^{(1)}_k,E^{(2)}_k\}$ of eigenvalues expected in some interval $(\omega_k,\omega_{k+1})$, and are therefore separated from the other eigenvalues by a quantity $\sim L^{-1}$. Thus, especially for small waveguide lengths, the energy separation of the complementary resonant states allows them to be coherently addressed and manipulated as a two-level system, provided the energy scale $g$ of the external coupling satisfies
\begin{equation}
    g \ll |E_{\nu} - \omega_{\bar{k}}| ,
\end{equation}
with $\bar{k}$ the closest integer to $\nu L/2d$. In this case, the subspace spanned by the two resonant states is a good candidate for a robust qubit, and coherent manipulation can occur by applying an external driving field (either time-dependent or time-independent, according to the performed tasks) to one or both emitters, while measurement in the computational basis can be performed by detecting the photon in the inner or outer region.

\textit{Conclusions.---} We have outlined the features of stable states in systems of one and two emitters coupled to a single transverse mode of a closed linear waveguide. In these eigenstates, the atomic and photonic excitations are dressed by interaction and hybridized with each other. In the two-emitter case, we have unearthed the existence of pairs of quasi-degenerate eigenstates, that represent the forerunners of the bound states in the continuum observed in unbounded geometries. Such a feature, which is absent in unbounded waveguides, opens the possibility to implement a qubit.  Future research will focus on specific entangled states in multi-emitter configurations and on the analysis of moving emitters~\cite{moving}, in which the interplay between internal and translational degrees of freedom can yield interesting effects.

\paragraph*{Acknowledgments}
PF and SP acknowledge support by MIUR via PRIN 2017 (Progetto di Ricerca di Interesse Nazionale), project QUSHIP (2017SRNBRK). ADG acknowledges the support of an ARC Future Fellowship (Grant No.\ FT160100357). PF and DL were partially supported by the Italian National Group of Mathematical Physics (GNFM-INdAM). 
PF, DL, FVP, and SP were partially supported by Istituto Nazionale di Fisica Nucleare (INFN) through the project ``QUANTUM'' and by Regione Puglia and  QuantERA ERA-NET Cofund in Quantum Technologies (GA No.\ 731473), project PACE-IN. 

\onecolumngrid
\appendix
\section{Eigenvalue equation}

\subsection{One emitter}

The state in the one-excitation sector has the form given in Eq.~(5) of the Letter:
\begin{equation}
\label{eq:state}
	\ket{\Psi} = a\ket{e}\otimes\ket{\mathrm{vac}}+\ket{g}\otimes\sum_{k}\xi_k b^\dagger_k \ket{\mathrm{vac}}
	= a \ket{e,\mathrm{vac}} + \sum_k \xi_k \ket{g,k } ,
\end{equation}	
and the action of the Hamiltonian $H= H_0 + H_{\mathrm{int}}$, given in Eqs.~(2)--(3) of the Letter, on the  basis vectors reads
\begin{align}
H \ket{e,\mathrm{vac}} & = \varepsilon \ket{e,\mathrm{vac}} +  \sum_{k}F_k \ket{g,k },
\nonumber\\
H \ket{g,k } & = \omega_k \ket{g,k } + F_k   \ket{e,\mathrm{vac}}.
\end{align}
Therefore, the eigenvalue equation $H \ket{\Psi}  = E \ket{\Psi} $ projected on the basis vectors gives
\begin{align}
&(E - \varepsilon) a + \sum_k F_k \xi_k =0, \nonumber\\
&(E- \omega_k) \xi_k + F_k  a =0.
\end{align}
By solving the second equation,
\begin{equation}
\xi_k = \frac{F_k}{E - \omega_k} a,
\end{equation}
and plugging it into the first one, we finally get
\begin{equation}\label{eq:eigenv}
\left(E - \varepsilon - \Sigma(E) \right) a =0,
\end{equation}
where 
\begin{equation}
\Sigma(E) = \sum_k \frac{F_k^2}{E-\omega_k}
\end{equation}
is the self-energy function of the model. This gives Eq.~(7) of the Letter.
The state normalization $|a|^2 + \sum_k |\xi|^k =1$ gives
\begin{equation}
|a|^2 \left(1+ \sum_k \frac{F_k^2}{(E - \omega_k)^2} \right) = |a|^2 \left(1 - \Sigma'(E)\right) =1,
\end{equation}
whence Eqs.~(8)--(9) of the Letter.

\subsection{Two emitters}

The computation is a straightforward generalization of the one-emitter case. The state in the one-excitation sector has the form given in Eq.~(12) of the Letter:
\begin{equation}
\label{eq:state_multi}
	\ket{\Psi} = \sum_{\alpha=1,2} a_\alpha \sigma^+_\alpha \ket{G} \otimes\ket{\mathrm{vac}} + \ket{G}\otimes \sum_k \xi_k b_k^\dagger\ket{\mathrm{vac}}
	= a_1 \ket{e_1,\mathrm{vac}} + a_2 \ket{e_2,\mathrm{vac}} + \sum_k \xi_k \ket{G,k } ,
\end{equation}	
and the action of the Hamiltonian $H= H_0 + H_{\mathrm{int}}$ on the  basis vectors reads
\begin{align}
H \ket{e_1,\mathrm{vac}} & = \varepsilon \ket{e_1,\mathrm{vac}} +  \sum_{k}F_k  \e^{\frac{-2\pi \ii k x_1}{L} }  \ket{G,k },
\nonumber\\
H \ket{e_2,\mathrm{vac}} & = \varepsilon \ket{e_2,\mathrm{vac}} +  \sum_{k}F_k  \e^{\frac{-2\pi \ii k x_2}{L} } \ket{G,k },
\nonumber\\
H \ket{G,k } & = \omega_k \ket{G,k } +  F_k  \e^{\frac{2\pi \ii k x_1}{L} }  \ket{e_1,\mathrm{vac}} + F_k  \e^{\frac{2\pi \ii k x_2}{L} } \ket{e_1,\mathrm{vac}}.
\end{align}
Therefore, the eigenvalue equation $H \ket{\Psi}  = E \ket{\Psi} $ projected on the basis vectors gives
\begin{align}
&(E - \varepsilon) a_1 + \sum_k  F_k \e^{\frac{2\pi \ii k x_1}{L} } \xi_k =0, \nonumber\\
&(E - \varepsilon) a_2 + \sum_k  F_k \e^{\frac{2\pi \ii k x_2}{L} }  \xi_k =0, \nonumber\\
&(E- \omega_k) \xi_k + F_k \e^{\frac{-2\pi \ii k x_1}{L} }  a_1 + F_k \e^{\frac{-2\pi \ii k x_2}{L} }  a_2  =0.
\end{align}
By solving the third equation,
\begin{equation}
\xi_k = \frac{F_k}{E - \omega_k} \left(\e^{\frac{-2\pi \ii k x_1}{L} }  a_1 +  \e^{\frac{-2\pi \ii k x_2}{L} }  a_2 \right),
\end{equation}
and plugging it into the first and the second one, we finally get
\begin{equation}
\left[(E - \varepsilon) \openone -\Sigma(E)\right] \bm{a}=0,
\end{equation}
where $\bm{a}= (a_1,a_2)^T$, and  $\Sigma(E) = \left(\Sigma_{j\ell}(E)\right)$, with
\begin{equation}
\Sigma_{j\ell}(E) = \sum_k \frac{F_k^2}{E-\omega_k} \e^{\frac{2\pi \ii k (x_j - x_\ell)}{L} },
\end{equation}
is the $2\times 2$ self-energy matrix of the model. This gives Eq.~(14) of the Letter. The energy of nontrivial solution must satisfy \begin{equation}
\det\left[(E-\varepsilon)\openone-\Sigma(E)\right]=0,
\end{equation} that is Eq.~(13) of the Letter.

\section{Calculation of the self-energy}
The self-energy of the model has matrix elements given by
\begin{equation}
\Sigma_{j\ell}(E)=  \sum_k  \frac{F^2_k}{E-\omega_k} \exp\left( \frac{2\pi \ii k (x_j-x_\ell)}{L} \right)=
\frac{\gamma}{L}\sum_{k=-\infty}^{\infty}  \frac{1}{\omega_k\left(E-\omega_k\right)} \exp\left( \frac{2\pi \ii k (x_j-x_\ell)}{L} \right),
\end{equation}
where 
\begin{equation}\label{eq:omegak}
	\omega_k=\sqrt{\left(\frac{2\pi k}{L}\right)^2+m^2}
\end{equation}
and $x_j,x_\ell$ are the positions of the emitters in the guide. The diagonal elements $\Sigma_{11}(E)=\Sigma_{22}(E)$ are nothing but the single-emitter self-energy function discussed in the first part of the Letter.
Notice that, because of the property $\omega_{-k}=\omega_k$, $\Sigma_{j\ell}(E)=\Sigma_{\ell j}(E)$; therefore, we can equivalently write
\begin{equation}
\Sigma_{j\ell}(E)= \frac{\gamma}{L}\sum_{k=-\infty}^{\infty}  \frac{1}{\omega_k\left(E-\omega_k\right)} \exp\left( \frac{2\pi \ii k d_{j\ell}}{L} \right),
\end{equation}
with $d_{j\ell}=|x_j-x_\ell|$. Also notice that the self-energy is invariant under the transformation
\begin{equation}
	d_{j\ell}\rightarrow L-d_{j\ell},
\end{equation}
as it must be since the physics of the system cannot depend on the orientation of the coordinate system.

To compute $\Sigma_{j\ell}(E)$ we will make use of the ``Herglotz trick'' of the cotangent, a beautiful argument which allows to determine the series by a complex integration~\cite{THE BOOK}. Consider the following function on the complex $\kappa$ plane:
\begin{equation}
f_{E,d}(\kappa)=\frac{\e^{\ii d\kappa}}{\sqrt{\kappa^2+m^2}\left(E- \sqrt{\kappa^2+m^2}\right)}\pi\left[\cot\left(\frac{\kappa L}{2}\right)-i\right],
\end{equation}
the complex square root to be interpreted in the sense of the principal value; we will fix $0<d<L$. For every real $E$, this is a meromorphic function in $\mathbb{C}\setminus[\pm \ii m,\pm \ii\infty)$ having
\begin{itemize}
	\item simple poles at $\kappa=\frac{2k\pi}{L}$ for all $k\in\mathbb{Z}$;
	\item two simple poles at $\kappa=\pm q(E)=\pm\sqrt{E^2-m^2}$, as far as $E> 0$;
	\item two branch cuts along $\pm \ii [ m, \infty)$.
\end{itemize}
Besides, it is exponentially bounded at $\infty$ in both half-planes: indeed, one gets
\begin{equation}
	\e^{\ii d\kappa}\left[\cot\left(\frac{\kappa L}{2}\right)-i\right]=2i\frac{\e^{id\kappa}}{\e^{\ii L\kappa}-1},
\end{equation}
whence its modulus for $\kappa=iy$, $y>0$, decays as $\e^{-dy}$, while for $\kappa=-iy$, $y>0$, it decays as $\e^{-(L-d)y}$.

\begin{figure}
	\includegraphics[width=0.9\linewidth]{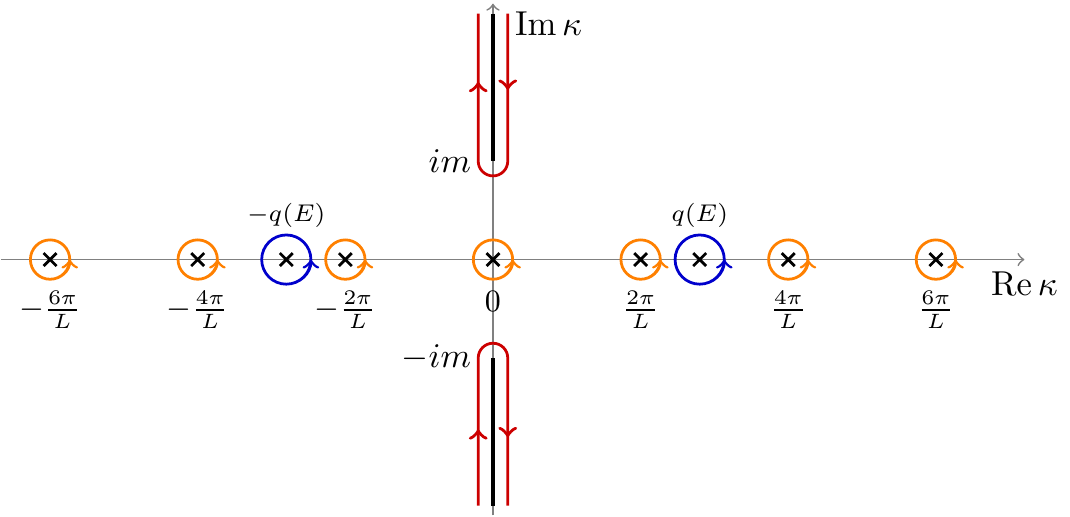}
	\label{fig:contour}
	\caption{Integration contour of the complex function $f_{E,d}(\kappa)$ in the complex plane. The integral over the path $\Gamma$ (the two red curves), here chosen in such a way to encircle the two branch cuts of the function, equals the sum of its residua, namely those corresponding to the points $\kappa=\frac{2\pi k}{L}$, $k\in\mathbb{Z}$ (orange small circles) plus the ones corresponding to the points $\kappa=\pm q(E)$, $q(E)=\sqrt{E^2-m^2}$ , as far as $E>0$ (blue large circles).}
\end{figure}

By the residue theorem, the integral of $f_{E,d}(\kappa)$ on any positively oriented contour $\Gamma$ (see Fig.~\ref{fig:contour}) which includes all its simple poles is  given by
\begin{equation}
	\frac{1}{2\pi \ii}\oint_{\Gamma}f_{E,d}(\kappa)\,\mathrm{d}\kappa=\sum_{k=-\infty}^{\infty}\Res_{f_{E,d}}\left(\frac{2\pi k}{L}\right)  +\Res_{f_{E,d}}\left(\sqrt{E^2-m^2}\right) \theta(E)+\Res_{f_{E,d}}\left(-\sqrt{E^2-m^2}\right) \theta(E),
\end{equation}
where $\theta(E)$ is the Heaviside step function.
An immediate calculation shows that
\begin{equation}
\Res_{f_{E,d}}\left(\frac{2\pi k}{L}\right)=\frac{2\pi}{L}\frac{1}{\omega_k(E-\omega_k)}\exp\left( \frac{2\pi \ii k d}{L} \right),
\end{equation}
while
\begin{equation}
\Res_{f_{E,d}}\left(\pm\sqrt{E^2-m^2}\right)=-\frac{\e^{\pm id\sqrt{E^2-m^2}}}{\sqrt{E^2-m^2}}\pi\left[\cot\left(\frac{L\sqrt{E^2-m^2}}{2}\right)\mp i\right],
\end{equation}
implying that
\begin{eqnarray}
	\frac{2\pi}{L}\sum_{k=-\infty}^{\infty}\frac{1}{\omega_k(\omega_k-E)}\exp\left( \frac{2\pi \ii k d}{L} \right)&=&2\pi\frac{\cot\left(\frac{L\sqrt{E^2-m^2}}{2}\right)\cos\left(d\sqrt{E^2-m^2}\right)+\sin\left(d\sqrt{E^2-m^2}\right)}{\sqrt{E^2-m^2}} \theta(E) 
	\nonumber\\
	& &+\frac{1}{2\pi\ii}\oint_{\Gamma}f_{E,d}(\kappa)\,\mathrm{d}\kappa.
\end{eqnarray}
Let us focus on the integral along the integration contour $\Gamma$; while the choice of $\Gamma$ is immaterial, it will be convenient to choose a contour which encircles both branch cuts of $f_{E,d}(\kappa)$ as in Fig.~\ref{fig:contour}, that is,
\begin{eqnarray}
	\oint_{\Gamma}f_{E,d}(\kappa)\,\mathrm{d}\kappa&=&\int_{im-0^+}^{i\infty-0^+}f_{E,d}(\kappa)\,\mathrm{d}\kappa-\int_{im+0^+}^{i\infty+0^+}f_{E,d}(\kappa)\,\mathrm{d}\kappa+\int_{-im+0^+}^{-i\infty+0^+}f_{E,d}(\kappa)\,\mathrm{d}\kappa-\int_{-im-0^+}^{-i\infty-0^+}f_{E,d}(\kappa)\,\mathrm{d}\kappa\nonumber\\
	&=&i\int_{m}^\infty\left(f_{E,d}(iy-0^+)-f_{E,d}(iy+0^+)\right)\,\mathrm{d}y-i\int_{m}^{\infty}\left(f_{E,d}(-iy+0^+)-f_{E,d}(-iy-0^+)\right)\,\mathrm{d}y\nonumber\\
	&=&-2\pi i\int_m^\infty\frac{\e^{-dy}}{\e^{-Ly}-1}\frac{1}{\sqrt{y^2-m^2}}\frac{2E}{E^2+y^2-m^2}\,\mathrm{d}y+2\pi i\int_m^\infty\frac{\e^{dy}}{\e^{Ly}-1}\frac{1}{\sqrt{y^2-m^2}}\frac{2E}{E^2+y^2-m^2}\,\mathrm{d}y\nonumber\\
	&=&4\pi i\int_m^\infty\frac{\e^{-dy}-\e^{-(L-d)y}}{1-\e^{-Ly}}\frac{1}{\sqrt{y^2-m^2}}\frac{E}{E^2+y^2-m^2}\,\mathrm{d}y\nonumber\\
	&=&4\pi i\int_m^\infty\frac{\cosh (dy)\coth\left(\frac{Ly}{2}\right)-\sinh (dy)}{\sqrt{y^2-m^2}}\frac{E}{E^2+y^2-m^2}\,\mathrm{d}y.
\end{eqnarray}
Notice that the latter contribution is invariant under the transformation $d\rightarrow L-d$. We finally get
\begin{equation}
	\Sigma_{j\ell}(E)=\gamma\frac{\cot\left(\frac{L\sqrt{E^2-m^2}}{2}\right)\cos\left(d_{j\ell}\sqrt{E^2-m^2}\right)+\sin\left(d_{j\ell}\sqrt{E^2-m^2}\right)}{\sqrt{E^2-m^2}} \theta(E)+\gamma \beta_{j-\ell}(E),
\end{equation}
where
\begin{equation}
	\beta_{j-\ell}(E)=\frac{1}{\pi}\int_m^\infty\frac{\cosh\left(d_{j\ell}y\right)\coth\left(\frac{Ly}{2}\right)-\sinh \left(d_{j\ell}y\right)}{\sqrt{y^2-m^2}}\frac{E}{E^2+y^2-m^2}\,\mathrm{d}y.
	\label{eq:betajl}
\end{equation}
This gives Eqs.~(15)--(16) of the Letter.
In particular, for $j=\ell$, one gets  $d_{j\ell}=0$ and the above expressions reduce to the single-emitter case:
\begin{equation}
	\Sigma(E)=\gamma\frac{\cot\left(\frac{L\sqrt{E^2-m^2}}{2}\right)}{\sqrt{E^2-m^2}} \theta(E)+\gamma \beta_{0}(E),
\end{equation}
with
\begin{equation}
	\beta_{0}(E)=\frac{1}{\pi}\int_m^\infty\frac{\coth\left(\frac{Ly}{2}\right)}{\sqrt{y^2-m^2}}\frac{E}{E^2+y^2-m^2}\,\mathrm{d}y.
\end{equation}
 This is the expression given in Eq.~(10) of the Letter.

Consider energies $E \geq 0$. As long as $d<L$, the integrand in~\eqref{eq:betajl} is a positive, monotonically decreasing function of $y$, since
\begin{equation}
\frac{\mathrm{d}}{\mathrm{d}y}\frac{\e^{-d_{j\ell}y}-\e^{-(L-d_{j\ell})y}}{1-\e^{-Ly}}=-\frac{1}{2}\biggl[(L-d_{j\ell})\cosh (d_{j\ell}y)+d_{j\ell}\cosh\left((L-d_{j\ell})y\right)\biggr]\operatorname{csch}\left(\frac{Ly}{2}\right)^2;
\end{equation}
consequently, we have
\begin{eqnarray}
	\beta_{j\ell}(E)&\leq&\frac{1}{\pi}\biggl[\cosh(md_{j\ell})\coth\left(\frac{mL}{2}\right)-\sinh(md_{j\ell})\biggr]\int_m^\infty\frac{1}{\sqrt{y^2-m^2}}\frac{E}{E^2+y^2-m^2}\,\mathrm{d}y\nonumber\\
	&=&\frac{1}{\pi}\biggl[\cosh(md_{j\ell})\coth\left(\frac{mL}{2}\right)-\sinh(md_{j\ell})\biggr]\frac{\operatorname{arcsinh}\left(\frac{\sqrt{E^2-m^2}}{m}\right)}{\sqrt{E^2-m^2}}\nonumber\\
	&=&\frac{1}{\pi}\biggl[\frac{\e^{-m(L-d_{j\ell})}+\e^{-md_{j\ell}}}{1-\e^{-mL}}\biggr]\frac{\operatorname{arcsinh}\left(\frac{\sqrt{E^2-m^2}}{m}\right)}{\sqrt{E^2-m^2}}.
\end{eqnarray}
In particular,
\begin{equation}
	\beta_{0}(E) \leq \frac{1}{\pi}\coth\left(\frac{mL}{2}\right)\frac{\operatorname{arcsinh}\left(\frac{\sqrt{E^2-m^2}}{m}\right)}{\sqrt{E^2-m^2}}
	\leq \frac{1}{\pi m}\coth\left(\frac{mL}{2}\right)
	\end{equation}
is positive and bounded for all positive energies. Moreover,
if \textit{both} $md\gg1$ and $m(L-d)\gg1$, i.e. the two emitters are sufficiently far away, $\beta_{1}(E)=\beta_{-1}(E)$ will be small and may be neglected,
\begin{equation}
|\beta_{1}(E)|=|\beta_{-1}(E)| \leq \frac{1}{\pi m}\left[\e^{-m(L-d)}+\e^{-md}\right];
\end{equation}
mathematically, the contribution of the upper and lower part of the contour in Fig.~\ref{fig:contour} yield, respectively, an $O(\e^{-md})$ and an $O(\e^{-m(L-d)})$  contribution to $\Sigma_{12}(E)$. 

\begin{figure}[h]
	\includegraphics[width=0.7\linewidth]{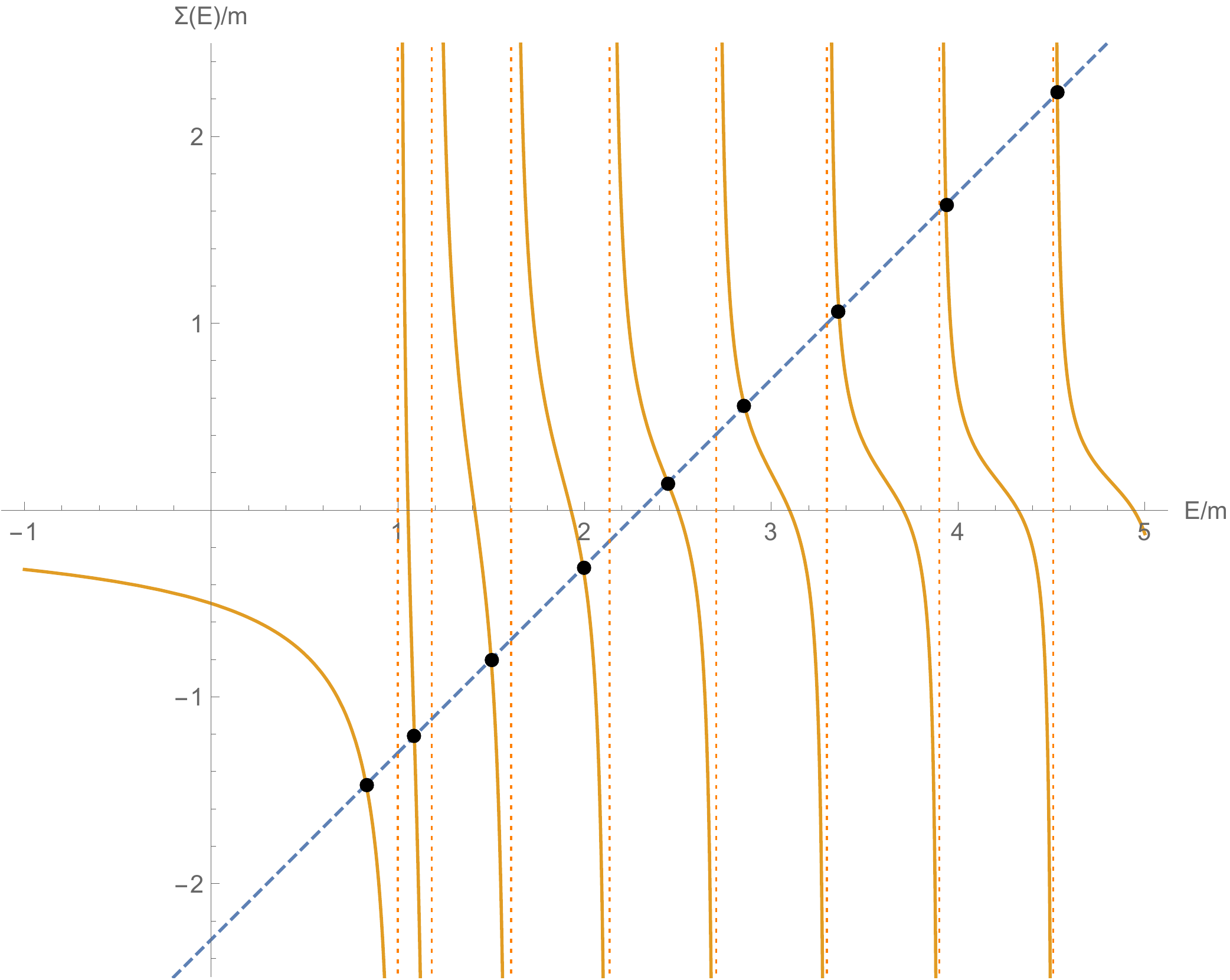}
	\caption{Solutions of Eq. \eqref{eq:eigenv} with $\varepsilon/m=2.3$ and $mL=10$. The orange solid curve corresponds to the self-energy, its asymptotes, represented as orange dotted vertical lines, corresponding to the values $\omega_k$, $k=0,\pm1,\pm2,\ldots$ in Eq. \eqref{eq:omegak}; the dashed blue line is the graph of $E-\varepsilon.$ The abscissa of each black dot is a solution of the eigenvalue equation.}
	\label{fig:intersection}
\end{figure}

Having computed the self-energy, in principle we can solve the eigenvalue problem for the model. In particular, in Fig. \ref{fig:intersection} we provide a graphical analysis of the solutions of the eigenvalue equation \eqref{eq:eigenv} for the single emitter in a waveguide for a suitable choice of the parameters; as discussed in the Letter, we have countably many eigenvalues above the mass threshold $E>m$, each being embedded between two consecutive eigenenergies $\omega_k,\omega_{k+1}$ of the uncoupled system. In particular, eigenvalues far from $\varepsilon$ will be close to the extrema of the interval $(\omega_k,\omega_{k+1})$, regardless the choice of the parameters, while eigenvalues close to $\varepsilon$ will lie somewhere in the middle and will be sensibly dependent on the parameters. Notice that the model will also exhibit a single eigenvalue below $m$, corresponding to an evanescent boson field concentrated around the emitter.

\twocolumngrid

\end{document}